\documentclass[8.5pt,twoside,twocolumn]{article}
\usepackage[super,sort&compress,comma]{natbib} 
\usepackage[version=3]{mhchem}
\usepackage{times,mathptmx}
\usepackage{sectsty}
\usepackage{balance} 

\usepackage{graphicx} 
\usepackage{amssymb}
\usepackage{lastpage}
\usepackage[format=plain,justification=raggedright,singlelinecheck=false,font=small,labelfont=bf,labelsep=space]{caption} 
\usepackage{fancyhdr}
\usepackage{natbib}
\usepackage{soul}

\def\tw{t_{\rm w}}
\def\tp{t_{\rm p}}
\def\tj{\Delta t_{\rm J}}

\usepackage{color}

\begin{document}

\thispagestyle{plain}
\fancypagestyle{plain}{
\renewcommand{\headrulewidth}{1pt}}
\renewcommand{\thefootnote}{\fnsymbol{footnote}}
\renewcommand\footnoterule{\vspace*{1pt}%
\hrule width 3.4in height 0.4pt \vspace*{5pt}} 
\setcounter{secnumdepth}{5}

\newcommand{\medio}[1]{\left\langle #1 \right\rangle} 

\def\<{\langle}
\def\>{\rangle}

\makeatletter 
\def\subsubsection{\@startsection{subsubsection}{3}{10pt}{-1.25ex plus -1ex minus -.1ex}{0ex plus 0ex}{\normalsize\bf}} 
\def\paragraph{\@startsection{paragraph}{4}{10pt}{-1.25ex plus -1ex minus -.1ex}{0ex plus 0ex}{\normalsize\textit}} 
\renewcommand\@biblabel[1]{#1}            
\renewcommand\@makefntext[1]%
{\noindent\makebox[0pt][r]{\@thefnmark\,}#1}
\makeatother 
\renewcommand{\figurename}{\small{Fig.}~}
\sectionfont{\large}
\subsectionfont{\normalsize} 

\fancyfoot{}
\fancyhead{}
\renewcommand{\headrulewidth}{1pt} 
\renewcommand{\footrulewidth}{1pt}
\setlength{\arrayrulewidth}{1pt}
\setlength{\columnsep}{6.5mm}
\setlength\bibsep{1pt}

\twocolumn[
  \begin{@twocolumnfalse}
\noindent\LARGE{\textbf{Particle jumps in structural glasses}
\vspace{0.6cm}

\noindent\large{
Massimo Pica Ciamarra\textit{$^{a,b}$}, Raffaele Pastore \textit{$^{b\dag}$} and Antonio Coniglio\textit{$^{b\ddag}$}} }\vspace{0.5cm}


\vspace{0.6cm}

\noindent \normalsize{
Particles in structural glasses rattle around temporary equilibrium positions,
that seldom change through a process which is much faster than the relaxation time,
known as particle jump. Since the relaxation of the system is due to the accumulation
of many such jumps, it could be possible to connect the single particle short time motion to the macroscopic
relaxation by understanding the features of the jump dynamics. Here we review recent results in this research direction, 
clarifying the features of particles jumps that have been understood and those
that are still under investigation, and examining the role of 
particle jumps in different theories of the glass transition.
}
\vspace{0.5cm}
 \end{@twocolumnfalse}
  ]

\footnotetext{\textit{$^{a}$~Division of Physics and Applied Physics, School of Physical and Mathematical Sciences, Nanyang Technological University,
Singapore}}  
\footnotetext{\textit{$^{b}$~CNR--SPIN, Dipartimento di Scienze Fisiche, University of Napoli Federico II, Italy}}  
\footnotetext{\dag~Corresponding e-mail address: pastore@na.infn.it}
\footnotetext{\ddag~ Corresponding e-mail address: coniglio@na.infn.it}

\section{Introduction\label{sec:introduction}}
Structural glasses, which are formed by many liquids cooled below their crystallization temperature,
provide an array of questions that is challenging researchers since many years.
In particular, as a liquid is supercooled one observes a striking difference
between the temperature dependence of the dynamical and of structural properties.
From the dynamical viewpoint one observes the relaxation time of these systems,
which is of the order picoseconds at the melting temperature,
to sharply increase on cooling. Conventionally, one considers
the supercooled liquid to fall out of equilibrium 
when the relaxation time reaches the value of
$100$s, and defines the corresponding temperature as the glass transition temperature, $T_g$.
This only slightly depends on the cooling rate.
At the glass transition temperature the dynamics is so slow that the liquid 
is in all respect an amorphous solid. As the temperature decreases
towards $T_g$ a class of supercooled liquids, known as fragile, also exhibits
qualitative changes in the relaxation dynamics. In particular,
above a temperature $T_x > T_g$ the dynamics satisfies the Stokes--Einstein (SE) relation,
the diffusivity being inversely proportional to the viscosity, while
below that temperature the SE relation is violated. 
Despite these reach dynamical features, 
the structural properties of these systems are found to be almost temperature independent. 
This is the `glass conundrum'.
From the thermodynamic viewpoint, a signature of the dynamical slowdown is only observed
close to the glass transition temperature, $T_g$, where the system falls out
of thermal equilibrium and the specific heat exhibits a jump.

The investigation of the single particle dynamics of supercooled liquids and structural
glasses offers a way to
reconcile the temperature dependence of dynamical and of structural quantities.
Indeed, on the one side the motion of a particle is affected by the local structure,
on the other side macroscopic relaxation occurs as particles eventually diffuse. 
Particle motion in supercooled liquids has been previously investigated,
and intermittency has emerged as a universal feature.
In supercooled liquids, particles rattle for a long time around a temporary equilibrium position, as in 
crystals, until at some time they jump and start rattling around a slightly different equilibrium position.
This intermittent motion is apparent in Fig.~\ref{fig:weeks_jumps}, that illustrates 
experimental results on the motion of colloidal particles in suspensions obtained by E.R. Weeks and D.A. Weitz\cite{weeks_subdiffusion_2002}.
This intermittent motion is referred as a cage--jump motion. It is usually described
by saying that the particles, caged by their neighbors, 
rattle in their cage for a while before jumping to a different cage,
even though this is a gross oversimplification of the actual physical
process taking place.

Here we review recent results on the cage--jump motion of supercooled liquids.
We have two goals. 
On the one side, we want to emphasize attempts 
to relate the cage--jump motion to the macroscopic relaxation of the system,
which would allow to connect the micro to macro scale. For instance, we will see that
it is possible to easily relate the diffusivity and the relaxation time
to features of the cage--jump motion, within the so called continuous time random
walk approximation. In addition, we will see that
there are attempts to relate other features of the dynamics of supercooled
liquids, such as dynamical heterogeneities, to the jumps.
On the other side, we want to clarify what is the role of
the single--particle jumps in different theories of the glass transition, including dynamical 
facilitation approaches, the mode coupling theory, the random first order theory
and elastic models of the glass transition. Reviewing previous
works that have investigated this question we suggest that 
a better understanding of the features of the cage--jump motion
could actually allow to reconcile some of these theories. In this
respect, the main open question results to be the degree of localization
of a jump.

We found it convenient to illustrate the key features of
the cage--jump motion by describing results obtained investigating a single system,
the prototypical Kob--Andersen Lennard--Jones three dimensional binary mixture~\cite{kob_testing_1995}. 
We thus refer to our own data, that mostly reproduce earlier results
obtained by others investigating different systems.
The paper is organized as follows. 
Sec.~\ref{sec:jumps} describes the jump process, and shows results clarifying
that jumps occur in cluster of few particles, which implies that a single particle jump
is the projection of a cooperative process on the trajectory of a particle.
Sec.~\ref{sec:compare} compares the clusters of jumping particles
with other dynamical particle clusters characterizing the relaxation of structural glasses,
namely dynamical and elastic heterogeneities. 
Sec.~\ref{sec:CTRW} considers successes 
and limitations of a jump--based kinetic description of the relaxation of glass formers, 
the continuous time random walk framework. 
Finally, in Sec.~\ref{sec:theory} we comment on the
role of these jumps in different theories of the glass transition. 
Future research directions and open problems are summarized in the conclusions.

\section{Jumps identification and properties~\label{sec:jumps}}
The introduction of jumps in the description of the dynamics of liquids
traces back to the classical hole model by Frenkel\cite{frenkel_y.i._kinetic_1946},
where liquids are depicted as sitting on distorted lattices in an attempt to describe
their short ranged order. 
The presence of a lattice suggests the introduction of
localized excitations associated to lattice vacancies and of an hopping dynamics in between lattice sites,
the jumps.
Indeed, early works\cite{chudley_neutron_1961} modeled the relaxation dynamics
of liquids as resulting from a sequence of jumps triggered by the presence of holes.
This dynamical process is clearly inspired by the hopping diffusion observed in crystals\cite{kittel_introduction_2005},
where (i) a jump involves a single particle and an hole,
(ii) the typical jump length is the interatomic spacing, which is roughly
temperature independent, (iii) a particle that jump changes its neighbors,
and (iv) jumps are uncorrelated, which implies that holes perform a random walk.

\begin{figure}[b!]
\begin{center}
\includegraphics*[scale=0.31]{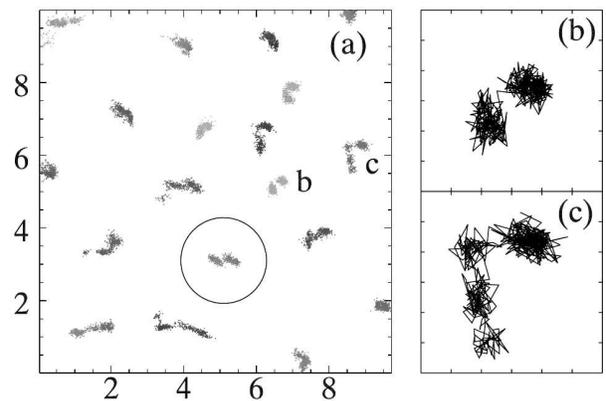}
\end{center}
\caption{\label{fig:weeks_jumps}
(a) Single particle cage--jump motion revealed from the trajectories
of particles in the bulk of an hard--sphere colloidal suspension, at a volume fraction $\phi = 0.52$. 
Axes are labeled in microns, and the circle illustrates the particle size. 
(b) and (c) are magnifications of two of the trajectories, with tick marks indicating spacings. 
Reprinted from Chemical Physics, 284, E.R. Weeks and D.A. Weitz,
Subdiffusion and the cage effect studied near the colloidal glass transition, Pages No. 361--367,
Copyright (2002), with permission from Elsevier.
}
\end{figure}

Subsequent works, however, clarified that 
the jumps identified in supercooled liquids differ
from those characterizing crystals in almost all of these features.
Perhaps the most important difference between the jumps observed in crystals
and those observed in liquids, is that the latter does not involve a single 
particle exchanging its position with an hole.
Rather, one observes a small number of close particles to perform a jump
at the same time\cite{candelier_spatiotemporal_2010,keys_excitations_2011, pastore_spatial_2015}; 
accordingly, a single--particle jump is the projection
of a cooperative motion on the trajectory of a single particle. 
The number of particles involved in such a cooperative rearrangement
is less than ten, and does not vary with 
temperature\cite{candelier_spatiotemporal_2010,keys_excitations_2011,pastore_spatial_2015}.
Because of this, jumps are considered localized excitations,
even though ascertaining their degree of localization is an important open issue,
as discussed in Sec.~\ref{sec:theory}.

A second important difference concerns the jump length and its temperature dependence.
Indeed, studies that have identified jumps via coarse--graining procedures allowing
for the determination of the jump length, found this to be
exponentially distributed, with an average size slightly decreasing on cooling\cite{helfferich_continuous-time_2014,pastore_dynamic_2015}.
The typical jump length, however, is always much smaller than the typical interparticle distance (see Fig.~\ref{fig:weeks_jumps}),
as opposed to the jumps occurring in crystals.
The decrease of the jump length with temperature makes the
identification of jumps problematic at low temperature. 
Furthermore, the small jump length implies that a particle
that performs a jumps does not usually change neighbors\cite{candelier_spatiotemporal_2010},
as in crystals.
Accordingly, local structural rearrangements associated with the change of one or
more neighbors\cite{souza_energy_2008,widmer-cooper_irreversible_2008,shiba_relationship_2012,lerner_statistical_2009}
are likely to occur through a sequence of jumps.

Successive jumps of a single particle could be either correlated, or uncorrelated; 
in an energy landscape perspective\cite{heuer_exploring_2008}, correlated and uncorrelated jumps 
correspond to transitions between inherent structures belonging to the same energy metabasin and to different energy metabasins, respectively.
Ascertaining whereas jumps are correlated or uncorrelated is a delicate task, and the
degree of correlation most probably depends on the operative definition of jumps. 
For instance, jumps can be identified\cite{hedges_decoupling_2007,michele_viscous_2001} by requiring a particle to move
more than a threshold $a$; in this case, they will be certainly identified as uncorrelated
if $a$ is large enough, as this procedure corresponds to the coarse graining of the trajectory 
of a particle that moves diffusively at long times.
Conversely, if jumps are identified via coarse graining procedures of particle trajectories that do not constraint the jump length,
then their correlation needs to be checked. 
A literature survey suggests that jumps may be correlated in the aging regime\cite{vollmayr-lee_single_2004},
and in polymer melts due to the topological constraint provided by the polymer chain\cite{helfferich_continuous-time_2014} (and references therein).
Conversely, in thermal equilibrium jumps appears to be uncorrelated\cite{pastore_dynamic_2015,pastore_spatial_2015}
if not in the deeply supercooled regime.

\begin{figure}[t!]
\begin{center}
\includegraphics*[scale=0.31]{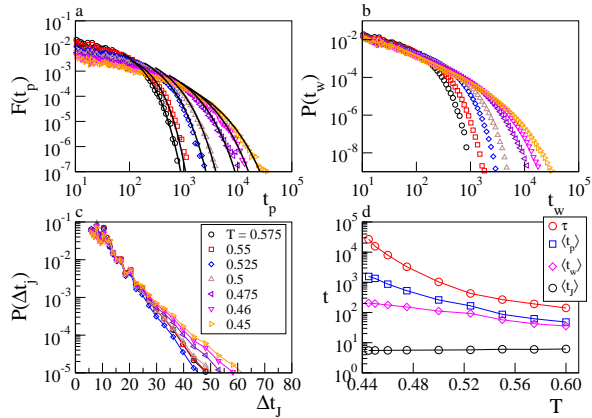}
\end{center}
\caption{\label{fig:timescales}
Probability distribution of the persistence waiting time (a), of the cage waiting time (b), and of the jump
duration (c), for the LJ Kob--Andersen mixture. Different curves refer to the different temperatures
as specified in the legend of panel (c). Panel (d) illustrates the temperature dependence of the
average values of these time scales. The figure also shows the temperature dependence of the relaxation time $\tau$ identified
from the decay of the self--scattering correlation function, $F(q,\tau) =1/e$, where $q$ 
is that corresponding to the maximum of the total structure factor.
Adapted form Ref.~\citenum{pastore_dynamic_2015}. Similar results have been observed in other atomistic
models of structural glasses\cite{hedges_decoupling_2007,candelier_spatiotemporal_2010} and gels\cite{chaudhuri_universal_2007}, 
and are commonly observed in facilitated lattice models (see Sec.~\ref{sec:theory}).}
\end{figure}

Three timescales characterize the jump motion.
The longest timescale is the average time a particle persists in its location before performing
its first jump, given an initial observation time. 
This is known as the persistence time, $\tp$.
The other timescales are the time $\tw$ a particle waits in a cage, as measured from the time the particle
entered that cage, and the jump duration, $\tj$. 
Fig.~\ref{fig:timescales} illustrates the probability distributions of these time scales
for the LJ Kob-Anderson mixture, as well as the temperature dependence of their average values and of the $\alpha$ relaxation time, $\tau$.
Similar results are found for other systems\cite{helfferich_glass_2015}.
We note that the persistent time and the waiting time
coincide at high temperature and do not differ much at low temperature, although they have a different temperature dependence.
Sec.~\ref{sec:CTRW} shows how these quantities allow to model the relaxation dynamics of the system within the continuous time random walk approximation,
and how the breakdown of the SE relation is related to these timescales.

\begin{figure}[t!]
\begin{center}
\includegraphics*[scale=0.31]{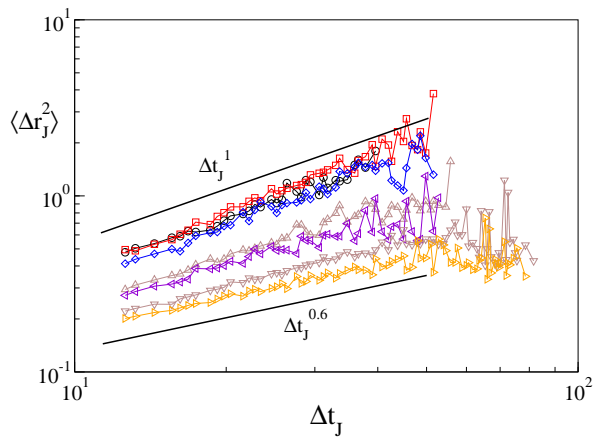}
\end{center}
\caption{\label{fig:subdiffusion}
Dependence of the squared jump length on the jump duration,
for the Kob--Andersen LJ mixture, for different temperatures as in Fig.~\ref{fig:timescales}.
The figure reveals that the jumps become increasingly subdiffusive on cooling.
Analogous results have been reported for different model systems\cite{pastore_spatial_2015}.
Due to the exponential distribution of the jump duration, illustrated in Fig.~\ref{fig:timescales}c,
the number of long--lived jumps is small, which is why data are noisy
at large values of the jump duration. Adapted from Ref.~\citenum{pastore_dynamic_2015}.}
\end{figure}

A detailed investigation of the dynamics of the jump process has not yet been accomplished.
A question of interest that has been clarified concerns the kind of motion performed by a particle while
jumping. For instance, one might expect that a particle jumps as its local environment
slightly change, giving rise to a net force acting on the particle.
If this is the case, particle motion during a jump should be superdiffusive. 
Conversely, one might expect that in order to jump a particle should overcome a free energy barrier\cite{schweizer_derivation_2005}, 
in which cases the motion should be subdiffusive. 
This question has been addressed by studying how the squared jump length scales with the jump duration.
Results obtained investigating the standard Kob-Anderson Lennard Jones mixture,
illustrated in Fig.~\ref{fig:subdiffusion}, and analogous results obtained investigating a system of Harmonic spheres\cite{pastore_spatial_2015},
show that jumps are diffusive at high temperature, and become subdiffusive on cooling.

We conclude this section by noticing that there is an aspect of the jump motion that has
not yet been properly investigated: its degree of localization.
Indeed, a small and temperature independent number of close particles is always seen to jump at the same time,
suggesting that jumps are localized over a typical and small temperature independent length scale.
However, it is also possible that these process requires the cooperation
of particles surrounding those that perform a jump.
If this is the case, the size of this surrounding region defines a 
cooperative length scale associated to the jumps. 
We name this length scale `cooperative jump length' for ease of reference.
Indeed, as discussed in Sec.~\ref{sec:theory}, ascertaining the existence
of the cooperative jump length and its temperature dependence might allow to contrast 
kinetic and thermodynamic theoretical descriptions of the glassy phenomenology.

\section{Connection with other localized excitations~\label{sec:compare}}
Particles performing a jump at the same time or in a short time interval
can be considered as a localized excitation. Here we consider what is the relation
between this excitation and the localized excitations identified by 
the dynamical and elastic heterogeneities.
Dynamical heterogeneities\cite{berthier_dynamical_2011} are clusters of particles with correlated motion,
that are commonly identified via the investigation of the time dependence
of the distribution of particle displacements, the van--Hove distribution function.
In supercooled liquids, the van Hove distribution has the Gaussian shape expected
in homogeneous systems at short times, when particle displacements 
reflect the velocity distribution, and at long times, when particles move diffusively.
At an intermediate timescale the van Hove distribution is not Gaussian, but has
long tails indicating that some particles moved much more than the average.
These particles can be identified using different approaches. A possible approach\cite{weeks_three-dimensional_2000}
is to identify them with the $5$\% of the particles with the largest displacement,
where this percentage is arbitrarily fixed. A different approach 
is to identify them\cite{kob_dynamical_1997,donati_stringlike_1998} with 
those particles whose displacement is larger than a threshold, which is fixed
by comparing the actual van-Hove function with that expected should the particles move diffusively.
Either case, particles contributing to the tail of the van Hove function
are found to be arranged in clusters, which we will refer to as dynamical heterogeneities.

While there are similarities between dynamical heterogeneities
and the cluster of particles that have performed a jump,
these should not be identified.
First, dynamical heterogeneities involve particle displacements
of the order of the interparticle distance. 
Indeed, they have been observed to lead to a string motion,
whereby a particle occupies the position previously occupied
by a different particle\cite{kob_dynamical_1997,donati_stringlike_1998}.
Conversely, the length of the jumps is much smaller.
For instance, for the usual LJ KA mixture, at a density $\rho = 1.2$,
the squared jump length $\< \Delta r_J^2 \>$ decreases from $0.25$,
at $T = 0.575$, to $0.1$, in the deeply supercooled regime at $T = 0.45$.
Assuming successive jumps to be uncorrelated, this implies that 
to move of a distance comparable to the
interparticle spacing, a particle must perform a number $n = \rho^{-2/d}/\< \Delta r_J^2 \>$
of jumps varying from $n \simeq 4$, at high temperature, to $n \simeq 11$, at low temperature.
Second, the typical size of dynamical heterogeneities 
increases as the dynamics slow down\cite{weeks_three-dimensional_2000}, at the point that these 
clusters have been associated to the growing cooperatively rearranging region\cite{kob_dynamical_1997,berthier_dynamical_2011},
while conversely the size of the clusters of jumping particles is constant.
Possibly, the size of DHs could be related to that of the cooperative jump length introduced in the previous section.
Finally, the typical time scale of dynamical hetergoeneities is larger than that of the jump dynamics.
All of these results suggest that dynamical heterogeneities result from a sequence of jumps.

The connection between particle jumps and dynamical heterogeneities is not fully understood.
An important observation is the presence of a facilitation mechanism, whereby a 
jump triggers jumps of nearby particles, in an avalanching process. This mechanism explains how DHs
emerge from a sequence of correlated jumps\cite{garrahan_geometrical_2002,candelier_spatiotemporal_2010,keys_excitations_2011,smessaert_distribution_2013,pastore_spatial_2015}. 
Facilitated models\cite{ritort_glassy_2003,chandler_dynamics_2010,pastore_dynamical_2011} offer a vivid picture of this process,
as they associate particles that jump to defects, and the avalanches to the diffusion of these defects.
However, the precise mechanism by which this facilitation mechanisms occurs
is not precisely understood. 
Indeed, one would expect the avalanche that develops through the facilitation 
process to spread like an infection, leading to a fast relaxation of the system. 
Conversely, avalanches are confined,
as they only grow until reaching a size of the order
of the dynamical heterogeneities.
Indeed, some particles appear immune to the facilitation process,
at least for a long time, as if requiring a large change to their
environment before being able to jump. In kinetically constrained models
(see Sec.~\ref{sec:theory}), these particles are those that belong to the core
of clusters of particles that cannot perform jumps due to kinetic constraints\cite{pastore_dynamical_2011}.
In the continuum, these particles are expected to be embedded in regions of high mechanical strength,
or with high structural order. The size of these regions has been observed to grow on cooling\cite{tanaka_critical-like_2010}.

Research in this direction should certainly consider the
presence of a third definition of cluster of localized particles,
that emerges from the analysis of the spatial properties
of the soft vibrational modes of amorphous materials. 
Indeed, in amorphous materials there are soft vibrational
modes that are localized: in the corresponding eigenvectors,
only few particles have a not negligible displacement,
and these few particles are localized in space. 
Earlier works called jumps the displacements associated to these modes,
and found them to comprise $10$ or more particles\cite{schober_low-frequency_1993,oligschleger_collective_1999}.
The soft modes are related to the spatial elastic heterogeneity of the system,
as they are localized in regions of the system characterized by small values of the local
shear modulus, or where the amplitude of vibration is 
large\cite{widmer-cooper_irreversible_2008,tsamados_local_2009,candelier_spatiotemporal_2010,mosayebi_soft_2014}.
These regions are known as `soft spots'~\cite{manning_vibrational_2011}.
Connections between dynamical heterogeneities
and soft spots, and between soft modes and jumping
particles, suggest an interplay between particle motion
and evolution of the local elastic properties.
Such an interplay, and particularly the evolution of the local elastic
properties, has not been yet carefully investigated. 
Recent works in these direction showed that soft
spots survive many elementary structural rearrangements, i.e. many jumps,
so many that their life time is correlated with the relaxation time of the system\cite{schoenholz_understanding_2014}. 
This might open the way to a coarse grained description of the relaxation
dynamics of supercooled liquids and structural glasses in term of soft spots.
The open question is that of determining the coarse--grained dynamical rules governing the evolution of the soft spots,
including their interaction, diffusion, annihilation and creation.

Finally, we consider that jumps have also been observed to play a role in the 
dynamics of amorphous glassy systems under shear. In these systems
the relaxation occurs through a sequence of relaxation events known
as shear transformation zones\cite{falk_dynamics_1998,bouchbinder_athermal_2007} (STZs).
STZs are extremely close to clusters of jumping particles.
Indeed, both involve few particles, both aggregate at long times
giving rise to DHs\cite{candelier_building_2009},
and both occur where the soft spots are localized\cite{manning_vibrational_2011}.
This analogy suggests that results obtained investigating STZs could also
hold for the cluster of jumping particles. This analogy is of interest
because the STZs, that occurs at zero temperature,
can be investigated in an energy landscape approach\cite{johnson_universal_2005}.
Research in this direction allowed to associate to the STZs a size related to
the number of particles involved in the rearrangement, of order\cite{johnson_universal_2005,pan_experimental_2008} $100$.
This number is sensibly larger than the number of particles undergoing large displacements in a jump, of order $10$.
Given the analogy between clusters of jumping particles and STZs, this finding supports the
idea that jumps do actually require cooperative rearrangements in
their surrounding regions, and therefore the existence of a jump cooperative correlation length.

\section{Continuous time random walk~\label{sec:CTRW}}
The continuous--time random walk formalism allows to make predictions concerning
the relaxation dynamics of supercooled liquids by assuming the absence of spatial and temporal correlations between the jumps.
Before describing these predictions, let us stress that the jumps are actually correlated,
even at high temperature. Indeed, uncorrelated jumps give rise
to particles with uncorrelated spatial positions, while conversely liquids have short ranged correlations.
Furthermore, at low temperature successive jumps of the {\it same} particle
are expected to be anticorrelated, as particles will jump back and 
forth between nearby positions before entering the diffusive regime.
Similarly, correlations between successive waiting times of a single particle are expected.
Despite these observations, it is interesting to consider what predictions
can be obtained for the relaxation dynamics when all of these correlations are neglected.
This strong approximation leads to the continuous time random walk (CTRW) description
of the relaxation dynamics of structural glasses.

The CTRW is a particle diffusion model, originally introduced by E.W. Montroll and G.H. Weiss\cite{montroll_random_1965},
that generalizes random walk processes by introducing
stochastic waiting times and stochastic jump lengths. 
With respect to other stochastic
diffusion models, e.g. Levy flights, the distribution of jump lengths is
assumed to have finite moments.
The CTRW proved useful to describe a variety of physical processes, including
electronic transport in disordered systems\cite{scher_anomalous_1975,nelson_continuous-time_1999}, 
diffusion in porous media or biological systems\cite{berkowitz_modeling_2006,wong_anomalous_2004,hofling_anomalous_2013},
blinking quantum dots\cite{tang_mechanisms_2005}, seismicity\cite{helmstetter_diffusion_2002}, spin glasses\cite{bouchaud_weak_1992}.
Notice that some of these processes are clearly out--of--thermal equilibrium. Out--equilibrium phenomena
concerning supercooled liquids that have been described in this framework include
aging\cite{helfferich_glass_2015}, and the diffusion of driven particles~\cite{schroer_anomalous_2013}.

These diverse systems share a similar dynamics consisting in subsequent jumps between energy (or free energy) minima.
The most important quantity fixing the time evolution is the distribution
of the waiting time in between jumps, $P(\tw)$, that in the case
of supercooled liquids has finite moments,
as all of the particles eventually jump. 
This distribution is exponential if jumps originate from
a Poissonian process. 
In supercooled liquids, this is only observed at high temperatures.
Conversely, as the temperature decreases, $P(\tw)$ develops
a power law regime at short $\tw$ and a stretched
exponential decay at large $\tw$, as in Fig.~\ref{fig:timescales}b.
This non--exponential behavior leads to a distinction between the probability
distribution of the persistence time, $F(\tp)$, which is the time
at which a particle perform its first jump as measured form an arbitrary initial observation time,
and the waiting time distribution, $P(\tw)$, the two distributions being
connected by the Feller relation\cite{lax_renewal_1977,tunaley_theory_1974},
$F(\tp) = \< \tw \>^{-1} \left(1-\int_0^{\tp} \psi(\tw) d\tw \right)$.

Within the continuous time random walk framework, it is straightforward to relate the 
diffusion coefficient, $D$, and the relaxation time at every wavevector $q = 2\pi/\lambda$ smaller than the average jump size,
to $\<\tw\>$, $\<\tp\>$ and to the average jump length $\<\Delta r_J\>$. One finds
\begin{eqnarray}
\label{eq:ctrw}
D &=& 6 \frac{\< \Delta r_J^2 \>}{\<\tw\>} \nonumber \\
\tau_\lambda &\propto& \<\tp\> + (m_\lambda-1)\<\tw\> + m_\lambda\<\tj\>, 
\end{eqnarray}
where $\tau_\lambda$ is estimated as the average time a particle needs to 
make $m_\lambda(T) = \lambda^2/\<\Delta r^2_J(T)\>$ jumps. In the above estimation
of $\tau_\lambda$ we have taken into account that jumps have a finite average duration $\<\tj\>$,
which is actually negligible at low temperatures as illustrated in Fig.~\ref{fig:timescales}.
There have also been attempts to describe the full time dependence of 
relaxation function\cite{berthier_length_2005} within the CTRW framework.

\begin{figure}[t!]
\begin{center}
\includegraphics*[scale=0.35]{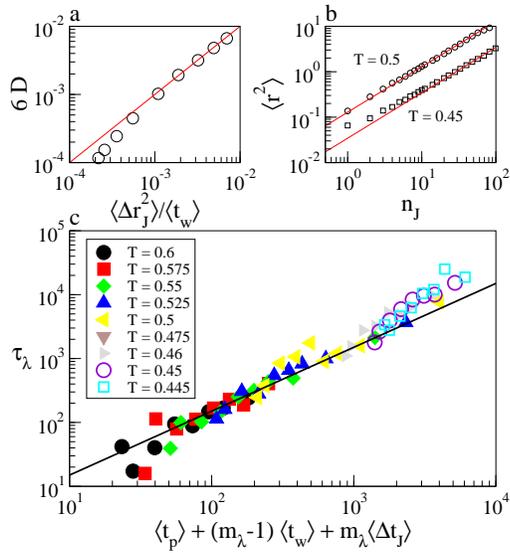}
\end{center}
\caption{\label{fig:ctrw}
Panels (a) and (c) show that the CTRW predictions (see text) reproduce measurements of the diffusion coefficient and of the relaxation
time at different wavectors of the KA Lennard--Jones mixture, if not at very low temperatures.
The failure at low temperature occurs as successive jumps of a same
particle becomes anticorrelated, as exemplified by the transient
subdiffuive dependence of the mean square displacement of the particles
on the number of jumps, illustrated in panel (b). 
}
\end{figure}

We have recently observed that these predictions work well for the KA LJ model systems\cite{pastore_dynamic_2015},
as illustrated in Fig.~\ref{fig:ctrw}, for a system of harmonic spheres\cite{pastore_cage-jump_2014}, 
as well in experiments of colloidal suspensions\cite{pastore_connecting_2014}.
This is an unexpected result given the expected presence of correlations between jumps
of a same particle and between jumps of different particles.
These correlations appear to be relevant only in the deeply supercooled regime,
where deviations from the CTRW predictions are observed.

The CTRW can also be related to DHs and to breakdown of the SE relation.
To clarify the relation with DHs, as measured for instance by the dynamical susceptibility,
we consider that DHs have both a temporal and a spatial contribution\cite{berthier_dynamical_2011}. 
The temporal one originates from fluctuations in the jumping time, while the spatial ones reflect the spatial correlation
of the jumping particles. The CTRW approach captures the temporal heterogeneity, as it allows for not exponential waiting time distributions.
This implies the temporary coexistence of particles that have performed many jumps, and of particles that have performed
few jumps\cite{pastore_dynamic_2015}. Spatial correlations are not captured as the CTRW approach nothing says about the spatial correlations of the particles
that have performed many jumps.
The temporal heterogeneities also allow to interpret the break down of the SE relation ($D\tau_\lambda = const.$).
Indeed, from Eq.~\ref{eq:ctrw} one understands that this breakdown 
is mainly due to the decoupling of the two timescales $\<\tw\>$ and $\<\tp\>$,
even though the temperature dependence of the squared jump length $\<\Delta r^2_J\>$
also plays a role.

\section{Jumps and theories of the glass transition\label{sec:theory}}
Since particle jumps are an ubiquitous feature of supercooled liquids, 
it is interesting to consider their role in theories
describing their slow dynamics and the glass transition.

\subsection{Jumps in kinetically constrained models} 
The cage-jump motion inspired the development of purely kinetic theories of the glass transition,
exemplified by kinetically constrained lattice models\cite{ritort_glassy_2003,chandler_dynamics_2010} (KCMs). 
In particle based models, particles sit on a lattice and move by hopping into nearest neighbors empty sites.
While this dynamics resembles the hopping motion of particles in crystals (see Sec.~\ref{sec:jumps}), 
these models are characterized by a kinetic constraint, according to which a jump is allowed provided the local environment
satisfies some condition. This kinetic constraint play a vital role, as it
implies that a jump is not a local process involving
a particle and an hole, but rather a cooperative process
involving few lattice sites. Because of this, a jump changes 
the local environment of nearby particles, and might allow or inhibit
their jumps. This process, termed kinetic facilitation\cite{ritort_glassy_2003}, 
creates correlations between subsequent jumps, and allows to rationalize the super Arrhenius temperature
dependence of the relaxation time\cite{garrahan_geometrical_2002,garrahan_coarse-grained_2003}.
The jumps introduced in kinetically constrained
models differ from those observed in supercooled liquids, as they imply particle
displacements of the order of the interparticle distance, the lattice cell size. 
While this could be a minor difference, it is also possible that these models overlook structural
reorganizations occurring at small length scales, that might introduce
a correlated dynamics mediated by an elastic coupling. 
To overcome this difficulty, one might consider the KCMs as a coarse grained description of an
actual physical system, in which case each site represents a spatial region, a jump inducing
the structural relaxation of that region.
Regardless of the physical interpretation, KCMs offers a simple and appealing picture of the dynamics slowdown of supercooled liquids,
but have two disadvantages. First, these models
are extremely schematic so that it is difficult to quantitatively relate them to different physical systems. 
Second, since kinetically constrained models have a trivial Hamiltonian, they are
by definition unable to describe the large specific heat jump occurring at the glass transition temperature.
Proposed remedies to this deficiency involve a better description\cite{biroli_are_2005,chandler_thermodynamics_2005}
of small scale jump processes. 
In this line of research, Odagaki\cite{odagaki_glass_1995,odagaki_microscopic_2000} first connected
the relaxation dynamics of supercooled liquids to their thermodynamic properties in a jump framework,
by assuming each jump to involve $n \propto 1/s_c(T)$ particles, where $s_c(T)$ is the excess entropy per atom of the supercooled liquid.

\subsection{Jumps in the free volume theory} The cage--jump motion has also been the primary source of inspiration of less
schematic theories of the glass transition, starting from the free volume theory by Cohen and Turnbul
\cite{cohen_molecular_1959,turnbull_free-volume_1961,turnbull_free-volume_1970}.
In this theory
the probability for a particle to escape from its cage only depends on its free volume.
Specifically, the theory assumes (a) that the total free volume $V_f$ of an $N$ particle
system is redistributed independently
among the particles, so that the single particle free volume probability distribution
is $p(v_f) = \< v_f \>^{-1}\exp(-v_f/\< v_f \>)$, where $\< v_f \> = V_f/N$, (b) 
that a particle is not able to jump out of its cage if its free volume is smaller than
a threshold, i.e. if $v_f < v_f^*$, (c) that the probability for a particle to
perform a jump does not depend on $v_f$, as long as $v_f > v_f^*$. From these assumptions,
one finds the escaping probability to be $P = \int_{v_f^*}^\infty p(v_f) d v_f$,
and can estimate the relaxation time $\tau$ to be 
$\tau = \tau_0/P = \tau_0 \exp(v_f^*/\<v_f\>)$, with $\tau_0$ a microscopic time.
Cohen and Turnbull estimated $\<v_f\> = A(T-T_0)$, and thus 
predicted a super-Arrhenius Vogel--Fulcher--Tamman (VFT) behavior for the relaxation time.

The free volume theory is extremely appealing due to its simplicity. We note that
it considers jumps to be collective processes, as in order to jump a particle
needs to have enough free volume, which is a property of the particle and of its immediate neighbors.
However, the theory also implicitly assumes that there is a small length scale process
that allows for the redistribution of free volume.
Because of this process, the free volume of all particles change in time; sooner or later,
all particles will have enough free volume to perform a jump. 
As the free volume theory does not model this free volume redistribution process,
it does not consider the presence of spatial correlations between jumping particles,
and says nothing as concern dynamical facilitation and dynamical heterogeneities.

\subsection{Jumps in the Mode Coupling Theory}
The idea that particles in supercooled liquids spend most of their time
rattling in the cage formed by their neighbors played a primary role in the development 
of the mode coupling theory\cite{bengtzelius_dynamics_1984,gotze_complex_2008,sjogren_kinetic_1980} (MCT)
of the glass transition.
The MCT, which is considered to provide a mean-field description of the dynamics of structural glasses,
is a first--principle theory that succeed in making specific predictions for the time evolution of correlation functions
of a liquid starting from its Lagrangian, using a projection operator approach (see Ref.~\citenum{reichman_mode-coupling_2005} for a review).
These predictions work well at high enough temperature, while they fail at lower temperatures. 
In particular, MCT predicts a power law divergence of
the relaxation time instead of the observed super-Arrhenius behavior, and does
not allow to rationalize the breakdown of the Stokes--Einstein relation.
To go beyond this mean field description fluctuations are taken into account
via the introduction of activated events, generally identified with hopping events.
In the mode coupling framework, this has been done either modifying the memory kernel
introducing an addition relaxation channels associated to these jumps\cite{das_fluctuating_1986,bhattacharyya_facilitation_2008},
as well as developing a dynamic free energy describing the escaping process\cite{schweizer_derivation_2005,rizzo_long-wavelength_2014,rizzo_nature_2015}.
While also called jumps, the activated processes invoked within this extended MCT theory have not 
a clear connection with the jumps discussed in this review. 
Indeed, particle jumps are observed in the slightly supercooled regime, where standard MCT works
and no additional relaxation channels are considered.
In addition, both approaches used to extend the MCT have not a clear relation with the jumps.
Indeed, when MCT is extended by modifying the memory kernel\cite{das_fluctuating_1986,bhattacharyya_facilitation_2008},
one actually simply introduces a new relaxation channel without making reference to its physical 
nature. That is, there is no reason to identify this new relaxation channel with the jumps described in the review.
Similarly, when the MCT is extended developing a dynamic free energy describing 
the escaping process
one finds the typical length of this escaping process to grow on cooling\cite{schweizer_derivation_2005},
or to have a non monotonic temperature dependence\cite{rizzo_long-wavelength_2014,rizzo_nature_2015}.
Conversely, all estimates of the single particle jump length suggest that this decreases on cooling.
The contrast between these two scenarios could be resolved whether jumps were found to 
be characterized by a growing cooperative jump correlation length, related to the size of the region
involved in the rearrangement rather than to the actual particle displacement, as discussed in Sec.~\ref{sec:jumps}.
The connection between a localized relaxation event and properties of the system on a large length scale has also been
recently suggested in free volume models\cite{coniglio_cell_2015}.

\subsection{Jumps in the Random first order theory}
The random first order theory (RFOT) of the glass transition
\cite{kirkpatrick_connections_1987,kirkpatrick_dynamics_1987,kirkpatrick_stable_1987,kirkpatrick_scaling_1989,cugliandolo_mean-field_1996,mezard_thermodynamics_1999}
introduced activated events to extend the mean filed description suggested 
by the analogy with $p$--spin models.
Simply put, in this picture a glassy system is seen as a collection of droplets, 
associated to the cooperatively rearranging region introduced by Adam and Gibbs\cite{adam_temperature_1965},
that continuously reconfigure
through activated processes\cite{biroli_random_2012}. 
The random first order theory models this process through a free energy for the reconfiguration of a droplet, 
introducing a bulk and a surface free energy in analogy with the free energy of nucleation of 
droplets in second order phase transitions. 
There is, however, not a consensus regarding the temperature and droplet size dependence
of this free energy.
Possibly, this question could be settled through a better understanding
of the dynamical process by which a droplet relax, that is expected to involve correlated single particle jumps.
In this respect, the contrast to be resolved is that existing from the length scale
characterizing the jumps, which is not temperature dependent, and the typical size
of the droplets, that conversely grows on cooling.
An appealing scenario involves, once again, the elastic properties of the system and the cooperative jump length,
as a jump may involve the deformation of a surrounding region. The size of this
region, associated to the droplet size, could be related
to point--to--set correlation lengths\cite{bouchaud_adam-gibbs-kirkpatrick-thirumalai-wolynes_2004,mezard_reconstruction_2006,montanari_rigorous_2006,biroli_thermodynamic_2008,kurchan_order_2011}.

\subsection{Jumps in Elastic models}
In Dyre's shoving model\cite{dyre_colloquium:_2006} the relaxation of supercooled liquids
is postulated to occur through localized structural rearrangements involving 
the overcoming of a free energy barrier. The model is termed
elastic, as the free energy barrier is associated to
the plateau shear elastic modulus of the system
\cite{hecksher_review_2015,larini_universal_2008}, $G_p$.
This association allows to predict that the relaxation time should scales as $\tau \propto \exp\left(G_p(T)/T \right)$,
in remarkable agreement with experimental data.
The dependence of the activation energy on the
shear elastic modulus suggests that a localized event actually involves the
deformation of the system on a large length scale. 
Since the model does not make specific assumptions regarding the features
of the relaxation events, it is hard to say whether these can be connected to the jumps.
However, in particular if the jumps are found to be characterized by a growing length scale,
their identification is certainly tempting. In this line of research, it would be interesting
to investigate how the energy barrier overcome during a jump is related to the 
plateau shear modulus.

\section{Conclusions}
In summary, we have reviewed results on the cage--jump motion
of supercooled liquids, and clarified that jumps
are processes involving the sensible displacement of a small group of particles. 
The size of these groups is
temperature independent, and each particle of the group move a distance
which is much smaller than the interparticle distance, and that decreases on cooling.
The mechanism by which these jumps accumulate leading to the relaxation
of the system, and to the glass phenomenology, is not fully understood,
even tough it certainly involves a facilitation process. 

More work is needed to clarify the connection between the jumps
and different theories of the glass transition. In this respect,
we notice that the extent to which jumps can be considered as localized
events has not yet throughly investigated. Indeed, it is certainly plausible
that in order for a jump to occur particles in a surrounding region
need cooperate. The size of this region would provide an estimate of a jump cooperative length. 
The investigation of the features of this
length, and particularly of its temperature dependence, could allow
to relate kinetic, elastic and thermodynamic theories of the glass transition.

\bibliography{CTRW}{}
\bibliographystyle{rsc}

\end{document}